# Magnetothermal cooling with a phase separated manganite

A. Rebello and R. Mahendiran[*]

Department of Physics and NUS Nanoscience & Nanotechnology Initiative

(NUSNNI), Faculty of Science, National University of Singapore,

2 Science Drive 3, 117542, Singapore

**Abstract**

We show that temperature of a current ($I$ = 20 mA) carrying manganite ($Nd_{0.5}Ca_{0.5}Mn_{0.93}Ni_{0.07}O_3$) in presence of a magnetic field ($H$) decreases abruptly as much as $\Delta T$ = 45 K (7 K) accompanied by a step like decrease in magnetoresistance at a critical value of $H$ when the base temperature is 40 K (100 K). The magnitude of $\Delta T$ and the position of magnetoresistance step decrease towards lower $H$ with decreasing amplitude of the current. We discuss possible origins of the current and magnetic- field driven temperature change which may find applications in magnetothermal refrigeration besides magnetocaloric effect.

---

[*] Corresponding author – phyrm@nus.edu.sg

The discovery of giant magnetocaloric effect (MCE) in Gd-Si-Ge alloy rejuvenated interest in search for new materials which can show a large change in isothermal magnetic entropy or adiabatic temperature change upon the application and removal of magnetic field.[1] MCE is the basis of the magnetic refrigeration technology which is considered to be more energy efficient and environmentally clean compared to the conventional vapor compression based technology. Besides a few other metallic alloys, colossal magnetoresistive manganites are also considered to be promising candidates for magnetic refrigeration due to the low cost of synthesis, chemical stability and high magnetic entropy change.[2] Another promising and simple technology for magnetic refrigeration is the Ettingshausen effect, which refers to the induction of a temperature gradient in a current carrying sample in a direction perpendicular to the direction of the applied magnetic field and dc current. In 1958, O'Brien *et al.* reported[3] a cooling of 0.25°C using a slab of single crystalline $B_{99}Pb_1$ alloy under $\mu_0 H$ = 0.1 T at room temperature. Later, Harman *et al.*[4] reported a temperature difference of 101 K between hot and cold ends in a wedge shaped Bi crystal under $\mu_0 H$ = 10.9 T. In 1994, Scholz *et al.*[5] constructed an Ettingshausen cooler with $Bi_{97}Sb_3$ alloy and attained a temperature difference of 42 K while maintaining the heat sink at $T$ = 160 K under $\mu_0 H$ = 0.75 T. As like the MCE, practical exploitation of the Ettingshausen effect awaits the discovery of a material showing a large temperature change around room temperature in relatively small magnetic fields offered by permanent magnets.

Although there are a few reports on the Nernst effect in manganites,[6] i.e., induction of a transverse voltage in a sample subjected to a temperature gradient perpendicular to *H*, possibility of magnetic cooling in presence of a dc current has been rarely studied.

The purpose of the present work is to investigate magnetothermal effect accompanying current driven magnetoresistance in the phase separated $Nd_{0.5}Ca_{0.5}Mn_{0.93}Ni_{0.07}O_3$ (NCMNi07). We show that increasing magnitude of the dc current in this samples transforms a smooth decrease of the resistivity ($\rho$) with $H$ into a step-like decrease (avalanche) at a critical value of $H$. Surprisingly, the avalanche in $\rho(H)$ is accompanied by an abrupt decrease in the temperature of the sample as much as $\Delta T \approx 40\text{-}45$ K. We also show that irreversibility of resistance observed at low dc currents upon $H$- field cycling is dramatically affected with increasing magnitude of the dc current.

We have measured the four probe resistance of a bar shaped polycrystalline NCMONi07 sample of dimensions $6\times3\times2$ mm$^3$, as a function of $H$ at different values of the dc current ($I$). During the magnetic field sweep, the surface temperature of the sample ($T_s$) was also measured simultaneously, with a Pt-100 sensor (3 x2x1.2 mm$^3$) glued to the top surface of the sample with a good thermal conductor (GE 7031varnish) between the voltage probes separated by 4 mm. The sample was mounted on a thin mica sheet which was thermally anchored to the sample puck with GE varnish. The base temperature of the sample, as recorded by a commercial cryostat is denoted by $T$ in the manuscript, unless otherwise stated. The dc current in the sample is perpendicular to the applied magnetic field. The details of the experimental set up were discussed in an earlier work.[7]

The main panel of Fig. 1 shows the temperature ($T$) dependence of the zero field dc resistivity ($\rho$) of the sample measured with $I$ = 100 µA. As like the 5 % Ni doped $Nd_{0.5}Ca_{0.5}MnO_3$ (Ref. 7), the present sample shows a semiconductor to metal transition while cooling and hysteresis while warming. The high value of $\rho$ at 10 K suggests that the sample is electrically and magnetically inhomogeneous due to coexistence of the ferromagnetic metallic (FM) and short-range charge-ordered (CO) antiferromagnetic insulating phases. The semiconductor to metal transition is caused by the percolation of FM clusters in the short range CO semiconducting matrix.[7] The inset shows the voltage-current (*V-I*) characteristics at 40 K on the left scale. The voltage shows a peak at a certain current value and exhibits a negative differential resistance behavior, accompanied by a significant increase in $T_s$ from 40 K to 60 K (right scale), which indicates the self heating effect in nonlinear electrical transport in manganites.[8]

Next, we show the influence of current amplitude on the *H*- field dependence of the $\rho$ at *T* = 40 K in Fig. 2. The left column shows the $\rho$-*H* isotherms measured for selected current strengths, (a) *I* = 100 µA, (b) 5 mA (c) 10 mA, and (d) 20 mA and the figures in the right column show the corresponding change in $T_s$. To erase the memory effect, *I* was reduced to zero after each field cycle and the sample was warmed up to 200 K, then cooled to 40 K. The complete field cycle consists of three consecutive sweeps: (1) 0 → +7 T, (2) +7 T → 0 → −7 T and (3) −7 T → 0 → +7 T. When *I* = 100 µA, during the sweep (1) (Fig. 2(a)), $\rho(H)$ decreases gradually until $\mu_0 H$ = 2.2 T is reached and then decreases more rapidly with further increasing *H*. Then, $\rho$ traces a different, low resistive path in the second sweep (2). In the third sweep (3), $\rho(H)$ still remains in the low resistive state and a small hysteresis is observed. This type of a large irreversibility and a persistent memory effect are found in other phase separated manganites also.[9,10] The $T_s$ increases gradually ($\Delta T \approx$ 1 K) with magnetic field in both the directions of the magnetic field (Fig. 2(e)) and it does not show any irreversibility and hysteresis unlike $\rho$. This suggests that the observed irreversibility in $\rho$ is intrinsic and is possibly related to the changes in the magnetic domain structure or domain wall nucleation upon reducing the field.

When $I = 5$ mA is applied, (Fig. 2(b)), $\rho(H)$ initially shows a sharp decrease near the origin followed by a gradual decrease with further increase in $H$ up to 1.65 T. However, $\rho$ shows a step- like decrease (avalanche), at a critical field $\mu_0 H_C = 1.7$ T. The $\rho$-$H$ curves in the sweeps (2) and (3) trace a low resistive path without any avalanche behavior. The sharp decrease in the resistivity near 0 T is accompanied by a sharp increase in $T_s$ from 40 K to 67 K (Fig. 2 (f)) and thereafter $T_s$ remains nearly constant at 67 K until field increases to 1.65 T. Then, $T_s$ also shows an avalanche effect, where in it abruptly drops to 41 K at $\mu_0 H_C = 1.7$ T, thereby coinciding with the avalanche in $\rho$. When $H$ is further increased from 2 T to 7 T, $T_s$ shows an increase less than 1 K. The temperature does not recover to a high value but remains low and shows a memory effect similar to the resistivity in the second and third sweeps. The field dependences of $\rho$ (Fig. 2(c)) and $T_s$ (Fig. 2(g)) at 10 mA are similar to 5 mA data except that the avalanche now occurs at a higher critical field, $\mu_0 H_C = 2.6$ T. A more spectacular effect appears in $\rho(H)$ when $I = 20$ mA (Fig. 2(d)). In the $H$- sweep (1) at $I = 20$ mA, $\rho$ shows the avalanche at $\mu_0 H_C = 3.8$ T followed by a gradual decrease at higher fields. When $H$ is reduced to zero, $\rho$ does not remain in the low resistive persistent state unlike what was observed for $I = 10$ mA and 5 mA. Instead, it shows the avalanche effect, but this time shows a step-like increase at $\mu_0 H = 0.8$ T. Upon reversing the field, $\rho(H)$ shows the avalanche effect again (high resistance to low resistance transition) at $\mu_0 H_C = -3.8$ T. A similar behavior of $\rho$ is observed in the third sweep with an abrupt increase from a low to a high resistive state at $\mu_0 H = -0.8$ T, followed by an abrupt decrease to a low resistive state again at $\mu_0 H_C = 3.8$ T. This, eventually results in a huge hysteresis far from the origin ($\mu_0 H = 0$ T). Note that $T_s$ follows a similar trend, which decreases abruptly from 86

K to 41 K when $\rho$ abruptly decreases at $\mu_0H_C = 3.8$ T during sweep (1) and then increases abruptly at $\mu_0H = 0.8$ T when the field is reduced to 0 T (Fig. 2(h)). Current-induced magnetoresistance step was also reported by Tokunaga et al.[10] in single crystals of Cr-doped $Nd_{0.5}Ca_{0.5}MnO_3$, however the temperature of the sample was not measured explicitly.

To illustrate the influence of the current strength on the evolution of the step-like decrease in $\rho(H)$, we show the $\rho(H)$ at $T = 50, 70, 100$ and $120$ K for $I = 20$ mA in Fig. 3(a). While $\rho(H)$ changes abruptly at a critical value of $H$ and exhibits a strong hysteresis with $H$-field cycling at $T = 50$ K, the critical field increases, abruptness in $\rho(H)$ decreases and the hysteresis vanishes with increasing temperature. The temperature of the sample exhibits a similar behavior (Fig. 3(b)). In Fig. 3(c), we have plotted the magnetic entropy ($\Delta S_m$) of the Ni-doped compound calculated from magnetization isotherms. The magnitude of $\Delta S_m$ is negative over the entire temperature range and shows a prominent maximum (4.2 J/kgK for $\Delta H = 7$ T) around $T_C$ (77 K) followed by a broad maximum of smaller magnitude (0.5 J/kgK) around the short-range charge ordering temperature ($T_{CO} = 252$ K). The magnitude of $\Delta S_m$ decreases with decreasing strength of the field. On the other hand, the undoped compound $Nd_{0.5}Ca_{0.5}MnO_3$ shows a negative peak in $\Delta S_m$ at a few K above $T_{CO}$, but changes into a positive peak at a few K below $T_{CO}$ (see the inset). A similar change in the sign of $\Delta S_m$ with temperature for a few CO manganites was reported earlier by Reis et al.[11] The negative $\Delta S_m$ observed over the entire temperature range in the present Ni-doped compound indicates that the lattice temperature of the sample will increase (decrease) upon adiabatic application (removal) of an external magnetic field similar to a normal ferromagnet. However, it alone will not account for the step-like decrease in the temperature observed at a critical value of the $H$- field in our compound. A detail study of MCE in these materials will be published later.

In order to confirm or rule out the possibility that the observed effect is caused by the relaxation of resistance, we have carried out the temporal dependence of the resistance and temperature of the sample. Figure 4(a) shows the evolution of resistance with time ($t$) at $T = 40$ K in a magnetic field, $\mu_0 H = 2.5$ T at different current strengths. When $I = 12.5$ mA or lower, the $R$ initially shows a small linear increase with the time and then becomes stable at $t \geq 30$ s. Surprisingly, when the current is increased by a small increment of 0.5 mA, i.e., when $I = 13$ mA, the $R$ increases linearly with the time up to $t = 200$ sec, but shows an abrupt increase around $t = t_{th} = 300$ sec. The $R$ shows a small peak at the end of the abrupt rise and then it quickly stabilizes. When the magnitude of the fixed current is further increased, the threshold time ($t_{th}$), where the abrupt increase in the resistance occurs, shifts towards a lower value of time. Note that the peak in the resistance become more prominent and it decreases in magnitude with increasing strength of the current. The temperature of the sample also undergoes an abrupt increase similar to the resistance as can be seen in Fig. 4(b). However, after the abrupt increase above $t_{th}$, $T_S$ remains constant unlike the resistance which shows a peak behavior. This study rules out the possibility that the step-like decrease in the resistance observed in the field sweep is caused by the relaxation of resistivity.

While the sudden increase in $T_s$ at the start of the $H$- sweep for $I \geq 5$ mA can be attributed due to the self Joule heating, the step-like decrease in $\rho$ accompanied by an abrupt decrease in $T_s$ at a critical field $H_C$ is unlikely due to the Joule heating effect. Note that ultrasharp multiple steps in magnetoresistance and magnetization at a critical $H_C$ below 5 K has been reported in some phase separated manganites.[9] Ghivelder et al.[12] found that the ultrasharp step in magnetoresistance and magnetization in phase separated $La_{0.225}Pr_{0.40}Ca_{0.375}MnO_3$ was accompanied by an abrupt decrease of the heat capacity and release of heat. The temperature of the sample increased abruptly from 2.5 K to 30 K at $H_C$ and then quickly relaxed to the initial temperature within a few milli seconds. The abrupt transition was suggested due to an intrinsic MCE, i.e., increase in the temperature of the sample during the field-induced metamagnetic transition. The zero field state at low temperature corresponds to a frozen state in which the CO and FM phases coexist. As the field is increased, some CO clusters with lower $H_C$, transforms into FM clusters and releases heat to the lattice in the process, which causes thermal instability of the neighboring CO clusters. In what follows as a chain reaction, the heat wave propagates through the volume of the sample within a few milli seconds.[13]

However, our results in the presence of a high dc current in the sample indicate an opposite behavior: both $\rho$ and temperature of the sample decrease abruptly at $H_C$. Why is it so? There are two possible scenarios. In the first scenario, it is caused by the decrease in the Joule heating accompanying the negative magnetoresistance. The conversion of the CO phase to FM phase at $H_C$ leads to lowering of the resistance (R) and hence the Joule heat dissipation ($P = I^2 \Delta R$, where $I$ is constant) decreases and in turn leads to the decrease in the temperature of the sample by $\Delta T = I^2 \Delta R / mC$, where $\Delta T$ is the change in temperature, $C$ is the heat capacity and $m$ (= 93 mg) is the mass of the sample. Upon reducing $H$ from the maximum value, the FM phase transforms back into a CO phase at a lower value of $H_C$, thus resulting in an abrupt increase of $\rho$. In turn, the power dissipation increases and hence the temperature of the sample also increases. At higher current strengths, the sample heats up more due to the Joule effect and hence the proportion of the CO phase increases relative to the FM phase. As a result, a higher value of $H$ is required to destroy the charge ordering and the critical field for metamagnetic transition shifts up with increasing current strengths. However, such a simple scenario does not explain why the temperature drop is so abrupt at a critical value of $H$. It does not also take into account of the changes in the destruction of CO state. Additionally, power dissipation based on the Ohm's law may not be completely valid for an electrically inhomogeneous system like our compound and joule heating working as a feed back to control the resistivity has to be considered.[14]

The second scenario is that the observed effect is related to the Ettingshausen effect, caused by heat transport carried by charge carriers deflected by the Lorentz force in high mobility semiconductors. If the current and the magnetic field are transverse to each other, the deflected electrons and holes recombine at the bottom of the sample, giving up their potential and kinetic energy to the lattice (sample heats up).[15] On the top side of the sample, electron-hole pairs must be created to correct the pair deficiency. This process requires absorption of energy from the lattice (sample cools) and the cooling effect continues until it is opposed by the Joule effect and by the heat conducted from the hot side. In our measurement, the temperature sensor is on the top surface of the sample and it possibly reflects the temperature of the whole sample itself, rather than only in the transverse direction to the current and magnetic field. Note that a significant temperature change of 30-40 K by the Ettingshausen cooling requires a current of about 10-80 A in Bi-Sb alloys[5] whereas our sample requires a current of 20 mA to induce a similar magnitude of the temperature change.

To summarize, $Nd_{0.5}Ca_{0.5}Mn_{0.93}Ni_{0.07}O_3$ exhibits a step like decrease in the resistivity, which is accompanied by an abrupt decrease in the temperature ($\Delta T$ = 40-45 K), at a critical magnetic field. The observed changes are neither caused by the Joule effect nor by the usual MCE. Further detail experimental investigations are necessary to ascertain the role of Ettingshausen effect in this magnetothermal phenomenon.

This work was supported by the office of DPRT, NUS (grant no. YIA-R144-000-197-123) and Ministry of Education, Singapore (grant no. ARF/Tier-1-R144-000-167-112).

**References**

[1] V. K. Pecharsky, and K. A. Gschneidner, Jr., Phys. Rev. Lett. **78**, 4494 (1997).

[2] M. -H. Phan, and S. -C. Yu, J. Magn. Magn. Mater. **308**, 325 (2007), and references therein.

[3] B. J. O'Brien and C. S. Wallace, J. Appl. Phys. **29**, 1010 (1958).

[4] T. C. Harman, J. M. Honig, S. Fischler, A. E. Paladino, and M. J. Button, Appl. Phys. Lett. **4**, 77 (1964).

[5] K. Scholz, P. Jandl, U. Birkholz, and Z. M. Dashevskii, J. Appl. Phys. **75,** 5406 (1994).

[6] R. Suryanarayanan, V. Gasumyants, and N. Ageev, Phys. Rev. B **59**, R9019 (1999); R. Suryanarayanan and V. Gasumyants, J. Phys. D: Appl. Phys. **35**, 2077 (2002).

[7] A. Rebello and R. Mahendiran, Solid State Commun. **149**, 673 (2009).

[8] A. S. Carneiro, R. F. Jardim, and F. C. Fonseca, Phys. Rev. B **73**, 012410 (2006); S. Imamori, M. Tokunaga, S. Hakuta, and T. Tamegai, Appl. Phys. Lett. **89**, 172508 (2006).

[9] R. Mahendiran, A. Maignan, S. Hébert, C. Martin, M. Hervieu, B. Raveau, J. F. Mitchell, and P. Schiffer, Phys. Rev. Lett. **89,** 286602 (2002).

[10] M. Tokunaga, H. Song, Y. Tokunaga, and T. Tamegai, Phys. Rev. Lett. **94**, 157203 (2005).

[11] M. S. Reis, V. S. Amaral, J. P. Araújo, P. B. Tavares, A. M. Gomes, and I. S. Oliveira, Phys. Rev. B **71**, 144413 (2005).

[12] L. Ghivelder, R. S. Freitas, M. G. das Virgens, M. A. Continentino, H. Martinho, L. Granja, M. Quintero, G. Leyva, P. Levy, and F. Parisi, Phys. Rev. B **69**, 214414 (2004).

[13] F. Macià, G. Abril, A. Hernández-Mínguez, J. M. Hernandez, J. Tejada, and F. Parisi, Phys. Rev. B **77**, 012403 (2008).

[14] A. V. Gurevich and R. G. Mints, Rev. Mod. Phys. **59**, 941 (1987); A. Palanisami, M. B. Weissman, and N. D. Mathur, Phys. Rev. B **71**, 094419 (2005).

[15] K. Seeger, *Semiconductor Physics*, 7th ed., pp. 95-101(Springer-Verlag, Berlin, 1999).

**Figures :**

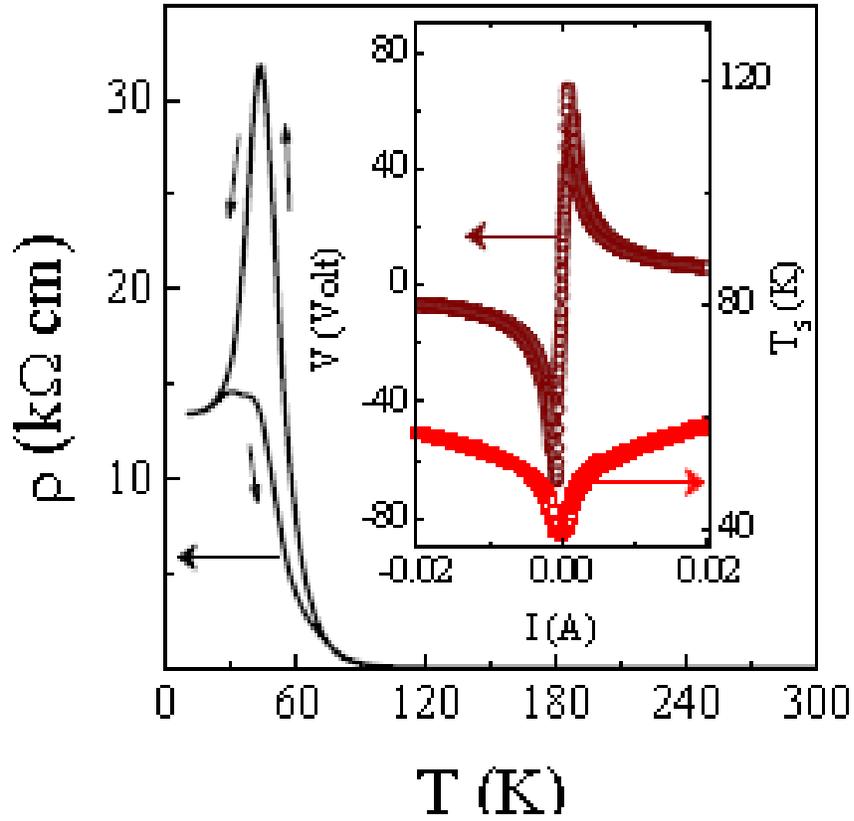

**FIG. 1**: (Color online) Main panel shows temperature dependence of the zero field *dc* resistivity ($\rho$) of the NCMONi07 and the inset shows the voltage-current (*V-I*) characteristics (on the left scale) at 40 K and the concomitant change in the surface temperature ($T_s$) (on the right scale).

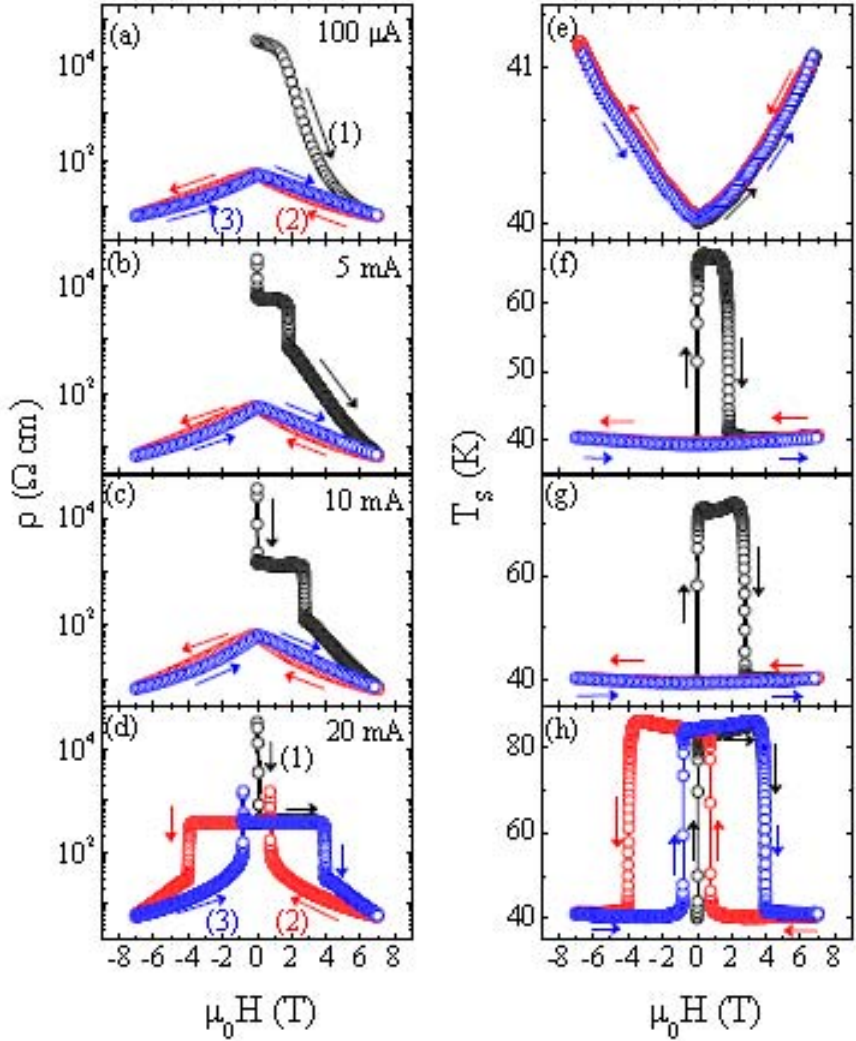

**FIG. 2**: (Color online) Left column shows the field dependence of resistance at $T = 40$ K for different current strengths and the right column shows the corresponding changes in the sample temperature ($T_s$).

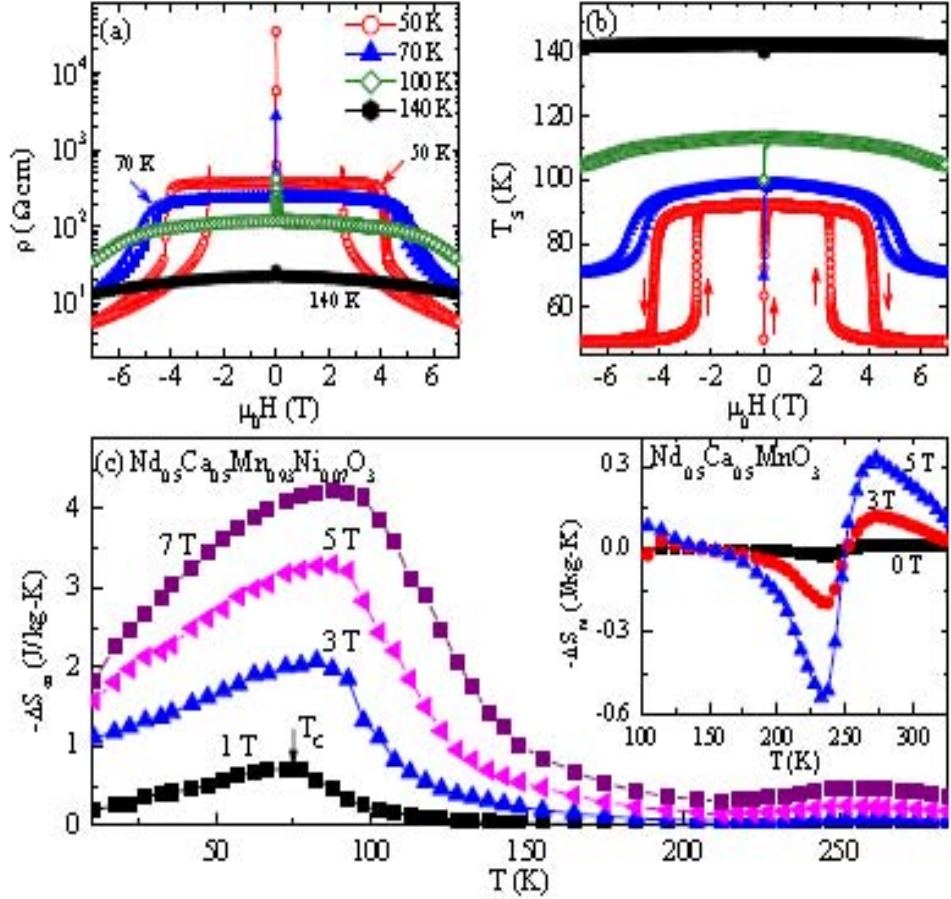

**FIG. 3**: (Color online) Magnetic field dependence of (a) $\rho$ and (b) $T_S$ at base temperatures indicated in the graph. (c) The main panel shows $-\Delta S_m$ vs. $T$ for different $\Delta H$ (1, 3, 5 and 7 T) for $Nd_{0.5}Ca_{0.5}Mn_{0.93}Ni_{0.07}O_3$. The inset shows $-\Delta S_m$ vs. $T$ for $Nd_{0.5}Ca_{0.5}MnO_3$.

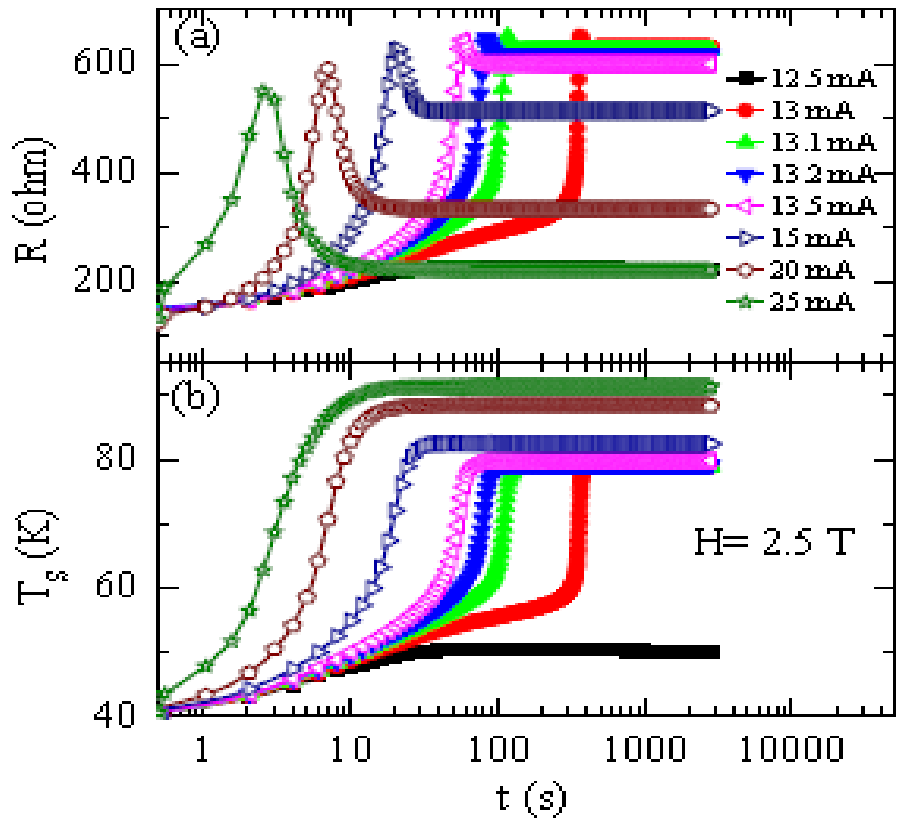

**FIG. 4**: (Color online) Time dependence of (a) resistance, $R$ and (b) temperature, $T_S$ at $\mu_0 H = 2.5$ T and $T = 40$ K.